\documentclass{article}
\usepackage{spconf,amsmath,graphicx}
\usepackage{multirow,booktabs,xurl}
\usepackage{hyperref}

\makeatletter
\newcommand{\thickhline}{%
    \noalign {\ifnum 0=`}\fi \hrule height 1pt
    \futurelet \reserved@a \@xhline
}
\makeatother

\title{WHISMA: A SPEECH-LLM TO PERFORM ZERO-SHOT SPOKEN LANGUAGE UNDERSTANDING}
\name{Mohan Li, Cong-Thanh Do, Simon Keizer, Youmna Farag, Svetlana Stoyanchev, Rama Doddipatla}
\address{Cambridge Research Laboratory, Toshiba Europe Ltd, Cambridge, UK}

\begin{document}
%
\maketitle
\begin{abstract}

Speech large language models (speech-LLMs) integrate speech and text-based foundation models to provide a unified framework for handling a wide range of downstream tasks. In this paper, we introduce WHISMA, a speech-LLM tailored for spoken language understanding (SLU) that demonstrates robust performance in various zero-shot settings. WHISMA combines the speech encoder from Whisper with the Llama-3 LLM, and is fine-tuned in a parameter-efficient manner on a comprehensive collection of SLU-related datasets. Our experiments show that WHISMA significantly improves the zero-shot slot filling performance on the SLURP benchmark, achieving a relative gain of 26.6\% compared to the current state-of-the-art model. Furthermore, to evaluate WHISMA's generalisation capabilities to unseen domains, we develop a new task-agnostic benchmark named SLU-GLUE. The evaluation results indicate that WHISMA outperforms an existing speech-LLM (Qwen-Audio) with a relative gain of 33.0\%.
\end{abstract}
\section{Introduction}
\label{sec:intro}

Traditional speech processing techniques typically depend on specialised models tailored to individual tasks. These models, trained with limited data and constrained architectures, often face difficulties in generalising to new domains and applications. However, recent advancements in speech foundation models (SFMs) \cite{baevski2020wav2vec,hsu2021hubert,chen2022wavlm} and large language models (LLMs) \cite{jiang2023mistral,achiam2023gpt,llama3modelcard} have fundamentally reshaped this approach, leading to the emergence of multi-functional end-to-end (E2E) speech-LLM systems \cite{wang2023blsp,tang2024salmonn,chu2023qwen,hu2024wavllm}. By utilising off-the-shelf SFMs and LLMs, the development of a speech-LLM system can be streamlined, often requiring only optimisation of adaptors and eliminating the need for extensive training data.

Similar to their text-based counterparts, speech-LLMs perform speech tasks through instruction following. Each task is defined by a textual prompt, which the LLM decoder processes alongside the speech embeddings produced by the SFM encoder. Conditioned on the speech input, the LLM then generates a response to fulfill the given instruction. Leveraging the vast knowledge embedded in the LLM decoder, speech-LLMs demonstrate emergent abilities not explicitly imparted during training \cite{tang2024salmonn}. Additionally, the prompt-driven approach enables speech-LLMs to address various speech classification tasks in a zero-shot manner. In this paradigm, users maintain the flexibility to incorporate different class labels into the prompt, rather than relying on fixed classification heads as used in conventional methods. These attributes notably enhance the potential of speech-LLMs to serve as a universal solution in the speech industry.

Previous studies have explored a wide range of speech-centric challenges within the speech-LLM framework. However, the evaluation of its spoken language understanding (SLU) capabilities, particularly in zero-shot scenarios, remains limited. While several existing speech-LLMs have showcased competence in common speech tasks like automatic speech recognition (ASR), speech-to-text translation (ST), and spoken question answering (SQA), their effectiveness in some key SLU tasks such as intent classification (IC) and slot filling (SF) has not been thoroughly examined. 

To address this gap, this paper introduces WHISMA, a speech-LLM system designed to enhance the zero-shot SLU performance across various domains. In this context, SLU refers to inferring semantics from spoken utterances. The proposed system employs Whisper \cite{radford2023robust} and Llama-3 \cite{llama3modelcard} models as the speech encoder and text decoder, respectively. These components remain fixed during training and are connected by a trainable modality aligner. Low-rank adaptation (LoRA) \cite{hu2022lora} is implemented on Llama-3 to accommodate speech modality inputs. WHISMA is fine-tuned using approximately 2000 hours of speech data, covering tasks of ASR, IC, SF, SQA, and spoken (query) instruction tuning (SQIT/SIT). We adopt a training strategy that enables the system to perform an auxiliary ASR step before SLU through speech chain-of-thought (SCoT) \cite{hu2024wavllm} or multi-round (MR) inference, while maintaining the E2E property of WHISMA.

To facilitate reproducing the proposed system, we ensure that all the training and test data used in this work is openly accessible. Specifically, we publish our self-developed Spoken-Alpaca dataset utilised for the SIT task. Derived from the text-based Alpaca instruction tuning corpus \cite{alpaca}, we conduct pre-processing on the input and instruction fields, and synthesise speech for them using an in-house text-to-speech (TTS) model. Moreover, we curate a new SLU benchmark, named SLU-GLUE, to assess speech-LLMs on task-agnostic data beyond IC and SF. Tasks within SLU-GLUE include sentiment analysis (SA), semantic equivalence recognition (SER), and spoken-textual entailment recognition (STER). This benchmark is also made publicly available.

The contributions of our work are summarised as follows:
\begin{itemize}
    \item We introduce WHISMA, a speech-LLM that incorporates cutting-edge speech (Whisper) and language (Llama-3) foundation models to support a diverse range of SLU tasks in zero-shot settings.
    \item Through evaluations on the SLU tasks seen in training, WHISMA demonstrates superior zero-shot performance compared to existing baselines on the SLURP \cite{bastianelli2020slurp}, FSC \cite{lugosch2019speech} and SmartLight \cite{saade2019spoken} benchmarks.
    \item To illustrate the robustness of WHISMA on unseen SLU tasks, we further evaluate the system on the curated SLU-GLUE benchmark\footnote{\url{https://drive.google.com/file/d/1mYwW-1ceU0R2dJEuaGv0iRaTb8-vmFfM/view?usp=drive_link}}, showing that it outperforms both the modular Whisper-Llama-3 system and an existing speech-LLM, Qwen-Audio \cite{chu2023qwen}.
    \item To mitigate over-fitting during WHISMA training, we develop the Spoken-Alpaca\footnote{\url{https://drive.google.com/file/d/1teWk0kUJ8u4hreb4JdVkCPqy2LnVFGuj/view?usp=drive_link}} dataset for speech-based instruction tuning, and show that it facilitates the model generalisation to unseen tasks.
\end{itemize}

\section{RELATED WORK}
\label{sec:format}

Motivated by the success of vision language models (VLMs) \cite{li2023blip,liu2023improved}, there has been a growing interest in extending LLMs with auditory capabilities. Initial attempts to incorporate speech perception into LLMs involve LLM-based ASR models \cite{li2023prompting,fathullah2024prompting}. Building upon this, Speech-Llama \cite{wu2023decoder} integrates both ST and ASR to the framework through multi-task learning. Further, SLM \cite{wang2023slm} enhances the system with speech instruction tuning. However, these models typically undergo fine-tuning with a limited set of tasks and lack generalised understanding abilities.

Recent advancements have introduced speech-LLMs with more diverse speech processing functions.  Notable examples include BLSP \cite{wang2023blsp}, LLaSM \cite{shu2023llasm}, SALMONN \cite{tang2024salmonn}, Qwen-Audio \cite{chu2023qwen}, and WavLLM \cite{hu2024wavllm}. Most of these systems utilise a variety of speech or audio resources to achieve cross-modal perception. Among them, zero-shot SLU evaluation results have been reported by BLSP for IC and SA tasks, SALMONN for a SF task, and WavLLM for a SQA task. In this work, we aim to conduct a comprehensive evaluation on these tasks using our proposed WHISMA system.

UniverSLU \cite{arora2024universlu} is the latest study that exclusively examines the SLU capabilities of the SFM, which fine-tunes the Whisper model with up to 17 SLU datasets encompassing both acoustic and semantic domain tasks. The zero-shot performance of UniverSLU, as reported, reveals that the system struggles to generalise to unseen datasets and tasks, which is the issue we would like to address in this work.

The most relevant research to this study is ZS-Whisper-SLU proposed by \cite{li2024prompting}, which investigates zero-shot IC and SF tasks within a QA-driven framework based on Whisper. Despite the promising performance achieved by the system, it is challenging to utilise ZS-Whisper-SLU for unseen tasks due to the limitation of the Whisper decoder. In this regard, we seek to bridge this gap by leveraging an LLM that possess broader knowledge than an SFM.

\section{METHOD}
\label{sec:method}

\begin{figure}[t]
\centering
  \centerline{\includegraphics[width=0.48\textwidth]{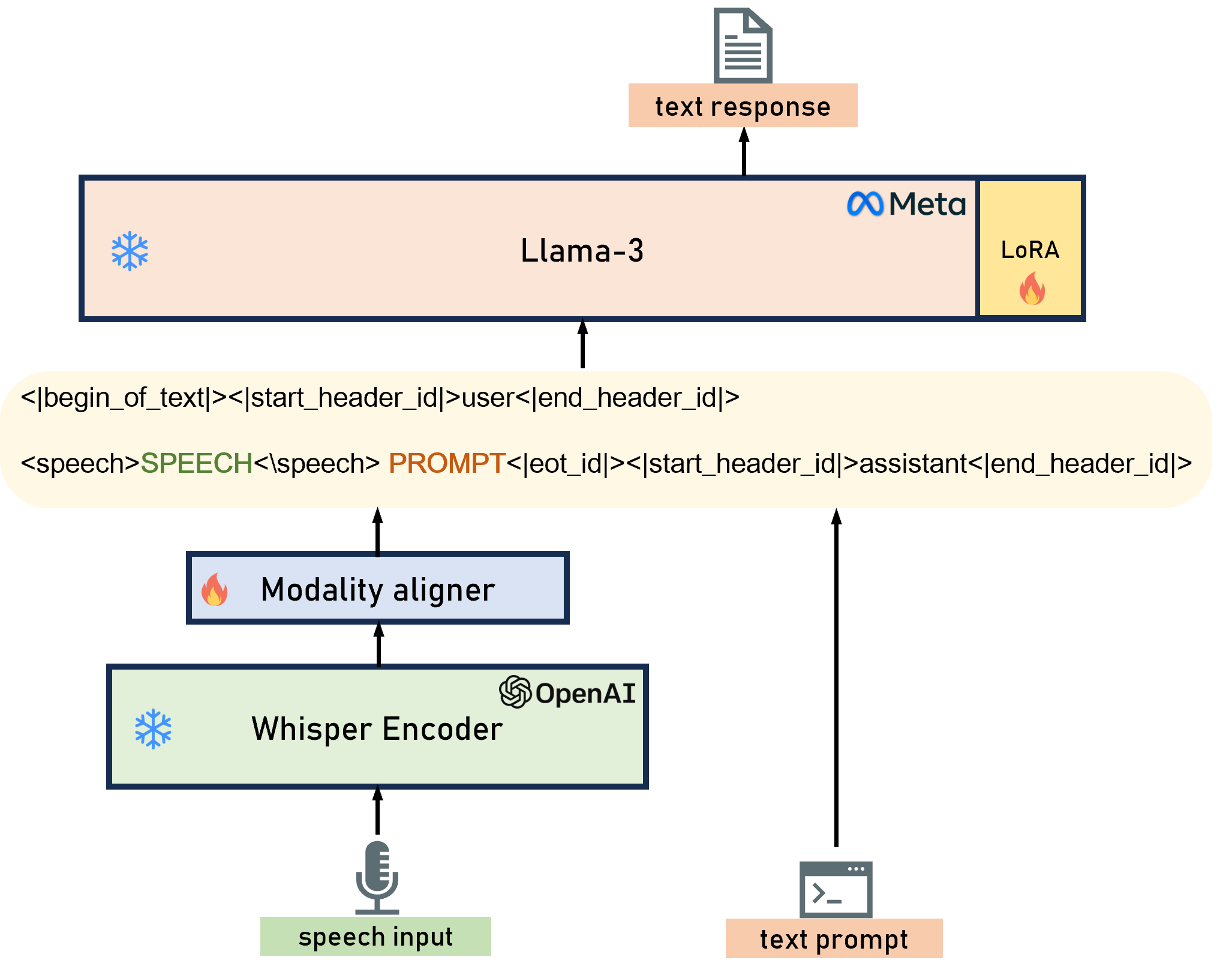}}
\caption{An overview of WHISMA model architecture.}
\label{fig:whisma}
\vspace{2mm}
\end{figure}

In this section, we present the architecture of the proposed WHISMA system and describe the fine-tuning speech tasks along with their corresponding datasets. Additionally, we introduce the training strategy that integrates ASR auxiliary into SLU tasks to enhance the reliability of LLM inference, while preserving WHISMA's end-to-end (E2E) nature.

\subsection{Model architecture}
\label{ssec:arch}

The model architecture of WHISMA is illustrated in Fig. \ref{fig:whisma}. We adopt the encoder from {\fontfamily{qcr}\selectfont Whisper-large-v2}, a 32-layer Transformer model with two convolutional neural network (CNN) down-samplers, as the speech encoder in our system. The resulting speech embeddings are fed to a modality aligner, which converts these embeddings to align with the input space of the LLM decoder. Our modality aligner is structured similarly to a previous system \cite{hu2024wavllm}, consisting of two CNN layers for further down-sampling, a bottleneck adaptor, and a linear output layer to match dimensions. 

WHISMA's decoder is implemented using {\fontfamily{qcr}\selectfont Llama-3-8B-Instruct}, one of the most promising open-sourced LLMs optimised for instruction following. This decoder processes both the aligned speech embeddings and a task-descriptive text prompt, generating a response to execute the given instruction. We integrate low-rank adaptation (LoRA) weights into the attention module of all 32 Transformer layers in Llama-3, enabling it to effectively handle speech modality inputs. During training, only the modality aligner and LoRA parameters are optimised.

\subsection{Training data}

\begin{table}[t]
\centering
\caption{Multi-task training dataset.}
\vspace{2mm}
\begin{tabular}{cccc}
\hline \thickhline
Task & Data Source & \#Hours & \#Samples \\  \thickhline \midrule
ASR & GigaSpeech (M) \cite{chen2021gigaspeech} & 1000 & 910k \\[3pt]
IC & SLURP (zero-shot) \cite{bastianelli2020slurp} & 80 & 94k \\[3pt]
SF & SLURP (zero-shot) \cite{bastianelli2020slurp} & 80 & 94k \\[3pt]
\multirow{2}{*}{SQA} & LibriSQA-partI \cite{zhao2023librisqa} & 360 & 104k \\[2pt] &LibriSQA-partII \cite{zhao2023librisqa} & 360 & 104k \\[3pt]
SQIT & Spoken-Alpaca & 30 & 29k \\[3pt]
SIT & Spoken-Alpaca & 14 & 14k \\[3pt] \hline \thickhline
\label{tab:data}
\end{tabular}%
\vspace{-0mm}
\end{table}

The fine-tuning of WHISMA is conducted with a multi-task learning approach. We compile a substantial training dataset that encompasses the following tasks: automatic speech recognition (ASR), intent classification (IC), slot filling (SF), spoken question answering (SQA), spoken query instruction tuning (SQIT), and spoken instruction tuning (SIT). Although ASR is not categorized as an SLU task, it significantly aids in aligning between the speech and text modalities. These tasks correspond to data from different sources, including the medium subset of GigaSpeech \cite{chen2021gigaspeech}, the zero-shot data split of SLURP \cite{bastianelli2020slurp}, LibriSQA-partI for open-ended QA, partII for multi-choice QA \cite{zhao2023librisqa}, and Spoken-Alpaca, developed as part of this research endeavor. The total duration of speech data amounts to approximately 2000 hours. Detailed information on this dataset is provided in Table \ref{tab:data}.

Although previous studies have employed Alpaca \cite{alpaca} with synthesised speech for instruction tuning \cite{hu2024wavllm,wang2023slm}, the associated data remains unpublished, which hinders the replication of their findings. To remedy this, we introduce Spoken-Alpaca and release it for public access. The original text-based Alpaca dataset comprises two types of examples: i) containing fields $\{instruction, input, output\}$, where $instruction$ describes the task, $input$ provides context, and $output$ is the expected system response; and ii) with fields $\{instruction, output\}$, where the $instruction$ stands alone as a query, not requiring contextual information. For the first type of examples, we generate speech for the $input$ field while retaining $instruction$ as the text prompt for the speech-LLM system. This configuration defines the task referred to as spoken instruction tuning (SIT). Concerning the second type, speech synthesis is applied to $instruction$, which the system receives without a text prompt. This task category is denoted as spoken query instruction tuning (SQIT). We develop an in-house variational inference with adversarial learning for end-to-end text-to-speech (VITS) model \cite{kim2021conditional} using the Librispeech \cite{panayotov2015librispeech} dataset featuring around 2200 speakers. Examples in Alpaca unsuitable for speech synthesis are filtered out, including those containing lengthy texts, mathematical equations, tables, etc. We also alter the wording of some instructions to make them speech-oriented. The diverse instructions within Spoken-Alpaca could significantly reduce the risk of over-fitting WHISMA to seen tasks with restricted prompt variations, such as ASR, IC and SF. Our experiments (Section \ref{ssec:result}) demonstrate that incorporating the dataset enables the proposed system to generalise more effectively to unseen tasks.

\begin{table*}[hbt!]
\centering
\caption{SLU-GLUE benchmark.}
\vspace{2mm}
\begin{tabular}{lclc}
\hline \thickhline
Task & Source & \multicolumn{1}{c}{Description} & \#Samples \\ \thickhline \midrule
SA & SST-2 & Classify the sentiment of \textbf{[SPEECH]} into positive or negative. & 2790 \\[3pt]
SER & QQP & Identify if the question in \textbf{[SPEECH]} is a paraphrase of the question in \textbf{[TEXT]}. & 3996 \\[3pt]
& QNLI  & Identify if the context in \textbf{[SPEECH]} contains the answer to the question in \textbf{[TEXT]}. & 2718 \\[2pt]
STER & RTE & Identify if the sentence in \textbf{[SPEECH]} entails the sentence in \textbf{[TEXT]}. & 2088 \\[2pt]
& SciTail & Identify if the premise in \textbf{[SPEECH]} supports the hypothesis in \textbf{[TEXT]}. & 2736 \\[2pt]
\hline \thickhline
\label{tab:slu-glue}
\end{tabular}%
\vspace{-0mm}
\end{table*}

\subsection{Training strategy}

The training examples are organised according to Llama-3's standard prompt template, as outlined in Fig. \ref{fig:whisma}. To enhance the robustness of WHISMA in handling diverse instructions during inference, we devise 10 distinct prompts for each task in ASR, IC, and SF. These prompts are randomly selected for each training example during fine-tuning. For the remaining tasks, we directly employ the provided question or instruction from the data as the text prompt. 

Unlike ZS-Whisper-SLU \cite{li2024prompting}, which tackles IC and SF through prompting the text decoder with individual questions for each intent class or slot type, WHISMA performs the tasks more efficiently by amalgamating candidate labels into a single prompt. Example prompts are: \textit{"Classify the intent of the spoken utterance into one of the following labels: [intent\_1], [intent\_2], ..., [intent\_N]"} for IC, and \textit{"Perform slot filling on the spoken utterance for the following slots: [slot\_1], [slot\_2], ..., [slot\_M]"} for SF. During training, we do not include all the candidate labels into the prompts but instead sample a varying number of them (including the ground-truth label) for each example. This approach introduces diversity to the prompts and helps prevent over-fitting. 

\begin{figure}[t]
\centering
  \centerline{\includegraphics[width=0.48\textwidth]{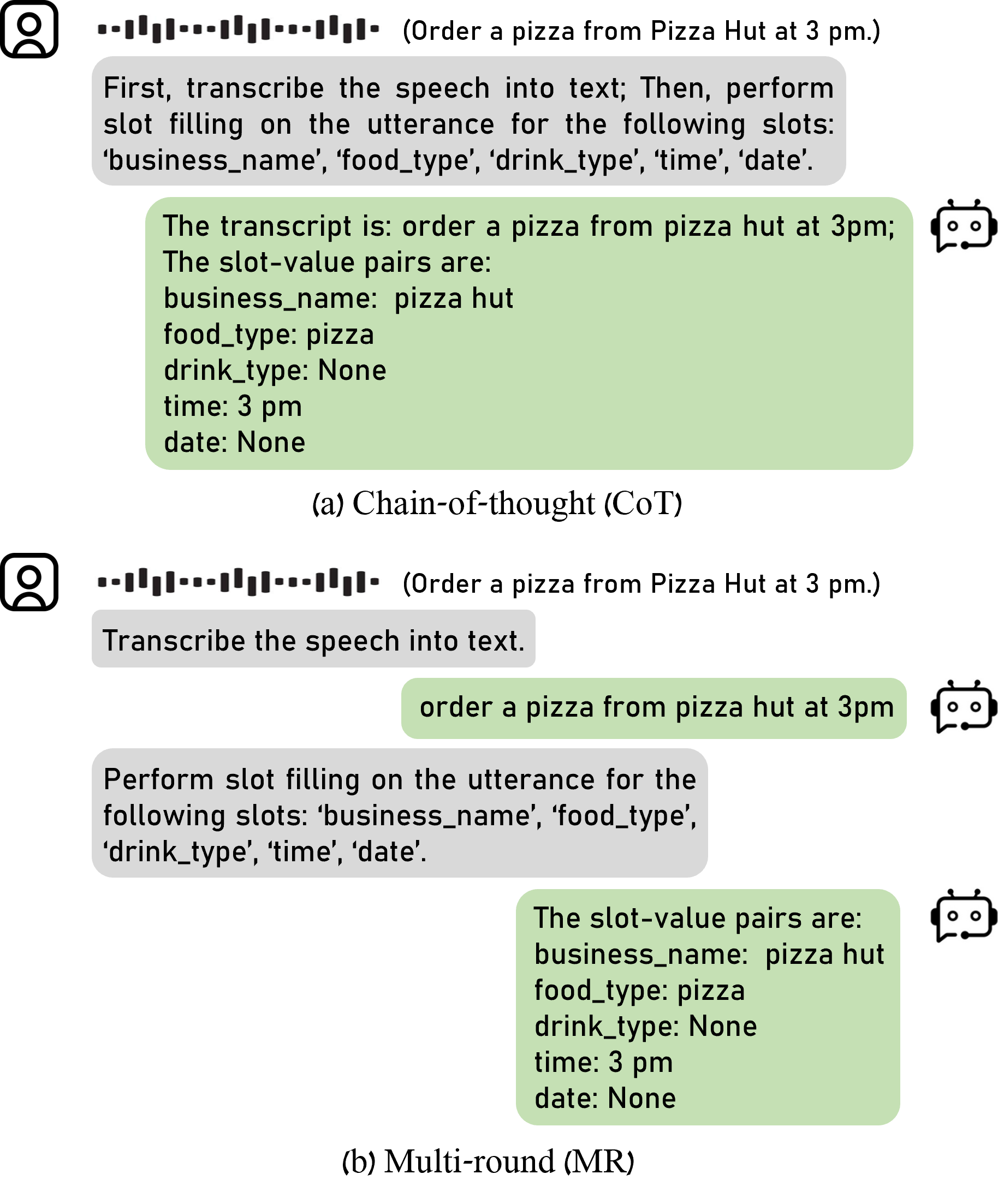}}
\caption{E2E Inference strategies integrating ASR to SLU.}
\label{fig:inf}
\vspace{-0mm}
\end{figure}

The primary challenge of E2E SLU lies in the direct extraction of semantic elements from speech, particularly for the SF task, which requires identifying entities explicitly mentioned in the user utterances. Previous studies indicate that ASR can significantly improve SLU performance \cite{li2023non}. We investigate this method for our proposed model. To distinguish WHISMA from modular ASR-LLM systems, speech transcription is performed as a part of the model inference, maintaining its E2E decoding property. We explore two strategies to integrate an auxiliary ASR step into SLU tasks:
\begin{itemize}
    \item Speech chain-of-thought (SCoT) \cite{hu2024wavllm}: The system processes a concatenation of the ASR and SLU prompts, executing the tasks one-shot in sequential order. During inference, the SLU response is conditioned on the previously generated speech transcript.
    \item Multi-round (MR): Inference is conducted in multiple dialogue rounds. The system is first instructed to produce a speech transcript, then based on the speech embeddings and the transcript (in the dialogue history), we prompt the system to execute the SLU task.
\end{itemize}

\noindent
An illustration of the above inference strategies is depicted in Fig. \ref{fig:inf}. In SLU tasks (excluding SQIT), each example is randomly assigned to one of the following training configurations: SLU-alone, SCoT-ASR-SLU, and MR-ASR-SLU. Despite the fact that LLMs are inherently capable of executing CoT and MR inference, our observation suggests that speech-LLMs do not develop such capabilities unless integrated into the fine-tuning process.

\section{EXPERIMENTS}
\label{sec:exp}

\subsection{Experimental setup}
\label{ssec:setup}

The speech encoder and text decoder of WHISMA are based on {\fontfamily{qcr}\selectfont Whisper-large-v2} and {\fontfamily{qcr}\selectfont Llama-3-8B-Instruct} models, respectively. The modality aligner adopts two 1-D CNN layers with a stride of 2, down-sampling the speech by a factor of 8 along with the Whisper encoder. This configuration produces 375 speech embeddings per input. The bottleneck dimension of the adaptor is set to 320. Within Llama-3, a light-weight LoRA scheme with a rank of 8 and an alpha of 16 is implemented. WHISMA is fine-tuned on 6 V6000 GPUs using a batch size of 12. To prevent over-fitting, the system undergoes only one epoch of training, using the AdamW optimiser with a constant learning rate of 0.0001.

\subsection{Evaluation tasks}
\label{ssec:eval}

We evaluate the proposed WHISMA model under three levels of zero-shot settings, ranging from easy to hard:
\begin{itemize}
    \item \textbf{seen-task-seen-corpus (STSC)}, targeting the SF task using the zero-shot split of the SLURP dataset \cite{bastianelli2020slurp}.
    \item \textbf{seen-task-unseen-corpus (STUC)}, focusing on IC and SF tasks within the test sets from FSC \cite{lugosch2019speech} and SmartLight \cite{saade2019spoken} benchmarks.
    \item \textbf{unseen-task-unseen-corpus (UTUC)}, representing the most challenging scenario, utilising a new SLU benchmark named SLU-GLUE.
\end{itemize}
For the STSC and STUC evaluations, we focus on data within the domain of in-home robot assistant. The same setups as described in \cite{li2024prompting} are followed to structure the test data. In the case of SLURP, a zero-shot data split is created by holding out 5 slot types: {\fontfamily{qcr}\selectfont\{podcast\_name, artist\_name, audiobook\_name, business\_name, radio\_name\}}, which results in a test set containing 18k utterances. Regarding FSC, the original {\fontfamily{qcr}\selectfont\{action, object, location\}} labels are reorganised into 15 intents and 2 slots. As for SmartLight, we employ the full close-field subset as the test set, comprising 6 intents and 3 slot types.

In addition to the task-oriented datasets mentioned above, we curate SLU-GLUE for the UTUC evaluation in our experiment. Derived from ASR-GLUE \cite{feng2021asr}, SLU-GLUE comprises three primary tasks: sentiment analysis (SA), semantic equivalence recognition (SER), and spoken-textual entailment recognition (STER). These tasks encompass five sub-tasks, namely: Stanford Sentiment Treebank (SST-2), Quora Question Pairs (QQP), Question-answering NLI (QNLI), Recognizing Textual Entailment (RTE), and SciTail. We exclude Semantic Textual Similarity Benchmark (STS-B) as it is not suitable for zero-shot evaluation. The evaluation metric for all the tasks is binary accuracy. The data domains of SLU-GLUE span from everyday language, such as movie reviews, to specialised scientific terminology. Speech samples in the dataset are recorded by six native speakers and are mixed with various levels of noise. Further detailed information of SLU-GLUE is provided in Table. \ref{tab:slu-glue}.

The performance of WHISMA is compared against multiple robust baselines, including state-of-the-art (SOTA) supervised models, SOTA zero-shot models, the modular ASR-LLM counterpart to WHISMA, and a well-established open-sourced speech-LLM called Qwen-Audio \cite{chu2023qwen}.

\subsection{Main results}
\label{ssec:result}

\begin{table}[t]
\centering
\caption{\textbf{STSC zero-shot evaluation.} WER (\%) and SF SLU-F1 (\%) on the zero-shot test set of SLURP.}
\vspace{2mm}
\begin{tabular}{lcc}
\hline \thickhline
Model & WER $\downarrow$ & SLU-F1 $\uparrow$ \\ \thickhline \midrule
\it{supervised} & & \\[2pt] \midrule
SNLU \cite{li2023towards} & 13.6  & 69.9 \\[2pt] \thickhline \midrule
\multicolumn{1}{l}{\it{zero-shot}} & & \\[2pt] \midrule
\multicolumn{1}{l}{ZS-Whisper-SLU \cite{li2024prompting}} &\multicolumn{1}{c}{8.3}  & \multicolumn{1}{c}{50.0} \\[3pt]
\multicolumn{1}{l}{WHISMA (proposed)} &\multicolumn{1}{c}{-} &\multicolumn{1}{c}{46.8}  \\[2pt]
\multicolumn{1}{l}{\quad + SCoT} &\multicolumn{1}{c}{11.9} &\multicolumn{1}{c}{63.1}  \\[2pt]
\multicolumn{1}{l}{\quad + MR} &\multicolumn{1}{c}{13.8} &\multicolumn{1}{c}{\textbf{63.3}}  \\[2pt]
\hline \thickhline
\label{tab:slurp}
\end{tabular}%
\vspace{-0mm}
\end{table}

\begin{table}[t!]
\centering
\caption{\textbf{STUC zero-shot evaluation.} WER (\%) and IC accuracy (IC Acc. \%) on the test set of FSC.}
\vspace{2mm}
\begin{tabular}{lcc}
\hline \thickhline
Model & WER $\downarrow$ & IC Acc. $\uparrow$ \\ \thickhline \midrule
\it{supervised} & & \\[2pt] \midrule
Finstreder \cite{bermuth2022finstreder} & -  & 99.7 \\[2pt] \thickhline \midrule
\it{zero-shot} & & \\[2pt] \midrule
BLSP \cite{wang2023blsp} & - & 77.5 \\[3pt]
\multicolumn{1}{l}{ZS-Whisper-SLU \cite{li2024prompting}} &\multicolumn{1}{c}{0.8}  & \multicolumn{1}{c}{95.0} \\[3pt]
\multicolumn{1}{l}{Whisper-Llama-3 (modular)} &\multicolumn{1}{c}{1.5}  & \multicolumn{1}{c}{89.3} \\[3pt]
\multicolumn{1}{l}{WHISMA (proposed)} &\multicolumn{1}{c}{-} &\multicolumn{1}{c}{90.9}  \\[2pt]
\multicolumn{1}{l}{\quad + SCoT} &\multicolumn{1}{c}{1.7} &\multicolumn{1}{c}{\textbf{97.3}}  \\[2pt]
\multicolumn{1}{l}{\quad + MR} &\multicolumn{1}{c}{2.7} &\multicolumn{1}{c}{\textbf{97.3}}  \\[2pt]
\hline \thickhline
\label{tab:fsc}
\end{tabular}%
\vspace{-0mm}
\end{table}

\begin{table}[t!]
\centering
\caption{\textbf{STUC zero-shot evaluation.} WER (\%), IC accuracy (IC Acc. \%),  SF SLU-F1 (\%), and perfect parsing (PP, \%) on the full set of SmartLight.}
\vspace{2mm}
\resizebox{\columnwidth}{!}{%
\begin{tabular}{lcccc}
\hline \thickhline
Model & WER $\downarrow$ & IC Acc. $\uparrow$ & SLU-F1 $\uparrow$ & PP $\uparrow$ \\ \thickhline \midrule
\it{supervised} & & & & \\[2pt] \midrule
Finstreder \cite{bermuth2022finstreder} & 6.1  & - & - & 88.0 \\[2pt] 
Whisper-TS \cite{arora2024universlu} & -  & 96.3 & - & - \\[2pt] \thickhline \midrule
\it{zero-shot} & & & & \\[2pt] \midrule
BLSP \cite{wang2023blsp} & - & 78.8 & - & - \\[3pt]
ZS-Whisper-SLU \cite{li2024prompting} & 2.7 & 91.6 & \textbf{90.9}  & 82.5 \\[3pt]
UniverSLU-14 \cite{arora2024universlu} & -  & 44.6 & - & - \\[3pt]
Whisper-Llama-3 & 4.0 & 94.3 & 82.4 & 75.7 \\
\quad(modular) \\[2pt]
WHISMA (proposed) & - & 81.7 & 87.8 & 68.9 \\[2pt]
\quad + SCoT & 4.5 & \textbf{95.9} & 90.2 & \textbf{82.5} \\[2pt]
\quad + MR & 5.5 & 94.4 & 90.7 & 81.6 \\[2pt]
\hline \thickhline
\label{tab:smartlight}
\end{tabular}%
}
\vspace{-0mm}
\end{table}

\begin{table*}[t]
\centering
\caption{\textbf{UTUC zero-shot evaluation.} Accuracy (\%) on SLU-GLUE sub-tasks.}
\vspace{2mm}
\begin{tabular}{lccccccccccc|c}
\hline \thickhline
Model & & SST-2 $\uparrow$ & & QQP $\uparrow$ & & QNLI $\uparrow$ & & RTE $\uparrow$ & & SciTail $\uparrow$ & & Avg. $\uparrow$ \\ \midrule
Qwen-Audio \cite{chu2023qwen} & & 53.4 & & 62.4 & & 58.8 & & 65.0 & & 54.5 & & 58.8 \\[2pt]
\quad + MR & & 52.9 & & 68.4 & & 52.4 & & 67.2 & & 56.3 & & 59.4 \\[3pt]
Whisper-Llama-3 (modular) & & 88.3 & & \textbf{74.9} & & 81.6 & & 76.7 & & 68.3 & & 78.0 \\[3pt]
WHISMA (proposed) & & 79.0 & & 65.1 & & 84.0 & & 72.8 & & 60.5 & & 72.3 \\[2pt]
\quad + SCoT & & 88.0 & & 65.1 & & 78.9 & & 73.7 & & \textbf{71.7} & & 75.5 \\[2pt]
\quad + MR & & \textbf{89.2} & & 74.2 & & \textbf{88.0} & & \textbf{78.0} & & 65.4 & & \textbf{79.0} \\ \midrule
WHISMA w/o Spoken-Alpaca & & 78.6 & & 45.9 & & 50.9 & & 61.2 & & 52.6 & & 57.8 \\[2pt]
\quad + SCoT & & 87.6 & & 69.0 & & 64.8 & & 74.6 & & 66.2 & & 72.4 \\[2pt]
\quad + MR & & 86.6 & & 52.6 & & 73.8 & & 70.0 & & 57.0 & & 68.0 \\[2pt]
\hline \thickhline
\label{tab:glue}
\end{tabular}%
\vspace{-0mm}
\end{table*}

The STSC evaluation results on the zero-shot test set of SLURP are presented in Table \ref{tab:slurp}, focusing on ASR word-error-rate (WER) and SF SLU-F1 score. To ensure a fair comparison, the modular Whisper-Llama-3 system is excluded as it has not been tuned on the SLURP dataset. For supervised model performance, we re-implement the spoken and natural language understanding (SNLU) system described in \cite{li2023towards} and conduct the same evaluation. Given the rich knowledge embedded in Llama-3 and the wide array of SLU datasets used in training, our proposed WHISMA system with either SCoT or MR inference significantly outperforms the existing SOTA model, ZS-Whisper-SLU, achieving a 26.6\% relative gain in SLU-F1. This comparison is equitable as ZS-whisper-SLU also conducts ASR prior to SF. The SCoT inference strategy achieves a lower WER compared to MR (11.9\% vs. 13.8\%), which we believe is because SCoT combines both tasks into a single prompt, enhancing ASR with the information in the SF instruction. WHISMA exhibits competitive SF performance (46.8\%) to ZS-Whisper-SLU even without utilising speech transcripts. However, it lags behind cases with SCoT or MR inference, which emphasises the importance of performing ASR for SF tasks.

The results of STUC evaluations on the FSC and SmartLight benchmarks are provided in Table \ref{tab:fsc} and \ref{tab:smartlight}, respectively. In the FSC benchmark, WHISMA+SCoT/MR obtains superior zero-shot performance compared to both the BLSP and ZS-Whisper-SLU baselines, showing relative improvements of 25.5\% and 2.4\% in IC accuracy. For the SmartLight benchmark, WHISMA+SCoT achieves SOTA performance in IC accuracy (95.9\%) compared to all the zero-shot baselines. It also delivers SF SLU-F1 (90.2\%) and perfect parsing (PP, 82.5\%) results comparable to ZS-Whisper-SLU. Similar performance is observed with MR inference. On both benchmarks, WHISMA without ASR still surpasses the modular Whisper-Llama-3 model in various scenarios, highlighting the reduced complexity of E2E speech-LLMs over the modular approach. Regarding the SCoT and MR inferences, due to the limited ASR data used in fine-tuning, WHISMA generally displays higher WER relative to ZS-Whisper-SLU and Whisper-Llama-3. However, the joint modelling of speech and text modalities in WHISMA+SCoT/MR compensates for this shortcoming, leading to improved SLU performance.

Table \ref {tab:glue} presents the UTUC evaluation results on the SLU-GLUE benchmark. Alongside the modular system, we include Qwen-Audio, the most comprehensive open-sourced speech-LLM (available at the time of this study), for comparison. Qwen-Audio serves as a relevant baseline to WHISMA, since its training data includes SLU tasks similar to those used in our work. Furthermore, the system supports MR inference. As shown in the table, WHISMA demonstrates promising generalisation capabilities to unseen tasks during training. Specifically, WHISMA+MR achieves the highest performance among all the speech-LLMs, with an averaged accuracy of 79.0\% across the five sub-tasks. This represents relative gains of 1.3\% and 33.0\% over Whisper-Llama-3 and Qwen-Audio+MR, respectively. Consistent with the observations in STSC and STUC settings, incorporating ASR into SLU substantially enhances WHISMA's performance on the challenging SLU-GLUE data, resulting in a relative gain of 9.3\% when MR inference is performed. In contrast, Qwen-Audio+MR only provides minimal improvements over the SLU-alone inference fashion. 

In Table \ref {tab:glue}, we further illustrate the impact of utilising the Spoken-Alpaca dataset into the fine-tuning process of WHISMA. One can see that excluding this dataset results in a notable performance decline for WHISMA, with a 20.1\% relative decrease in averaged accuracy when ASR is omitted. Similar degradation is observed with SCoT and MR inference, suffering relative reductions of 4.1\% and 13.9\% in averaged accuracy, respectively. Therefore, although Spoken-Alpaca contributes only a small fraction (2.2\% of total duration) of the entire training data, it plays a crucial role in strengthening the generalisability of speech-LLMs.

\section{Conclusion}
\label{sec:con}

In this paper, we present WHISMA, a speech-LLM system that excels in various zero-shot SLU tasks. WHISMA integrates a Whisper-based speech encoder with a Llama-3 text decoder, and is fine-tuned on a diverse set of SLU tasks using a modality aligner and LoRA adaptors. Additionally, we enable the system to perform an auxiliary ASR step before SLU through SCoT or MR inference strategies. Comprehensive zero-shot evaluations demonstrate WHISMA's ability to achieve SOTA performance on several common SLU benchmarks and, more importantly, to generalise to tasks not encountered during training. To ensure the reproducibility of our results, we release the Spoken-Alpaca and SLU-GLUE datasets utilised in our experiments for public access.


\bibliographystyle{IEEEbib}
\bibliography{refs}

\end{document}